\documentclass[a4paper,11pt]{article}
\pdfoutput=0 

\usepackage{jheppub} 

\usepackage[T1]{fontenc} 
\usepackage{multirow} 
\usepackage{color,ulem}
\usepackage{amsfonts} 
\usepackage{amssymb}
\usepackage{amsmath}  
\usepackage{amsbsy}  
\usepackage{epsfig} 
\usepackage{wasysym}
\usepackage{graphicx}

\def\beq{\begin{equation}}
\def\eeq{\end{equation}}
\def\bea{\begin{eqnarray}}
\def\eea{\end{eqnarray}}
\def\lsim{\mathrel{\raisebox{-.6ex}{$\stackrel{\textstyle<}{\sim}$}}}

\newcommand{\dsl}[1]{\not \hspace{-0.7mm}#1}

\newcommand{\eg}{{\it e.g.}}

\def\@citex[#1]#2{\if@filesw\immediate\write\@auxout{\string\citation{#2}}\fi
 \def\@citea{}\@cite{\@for\@citeb:=#2\do
    {\@citea\def\@citea{,\penalty\@m}\@ifundefined
       {b@\@citeb}{{\bf ?}\@warning
       {Citation `\@citeb' on page \thepage \space undefined}}%
\hbox{\csname b@\@citeb\endcsname}}}{#1}}
\def\citer{\@ifnextchar
[{\@tempswatrue\@citexr}{\@tempswafalse\@citexr[]}}

\def\@citexr[#1]#2{\if@filesw\immediate\write\@auxout{\string\citation{#2}}\fi
  \def\@citea{}\@cite{\@for\@citeb:=#2\do
    {\@citea\def\@citea{--\penalty\@m}\@ifundefined
       {b@\@citeb}{{\bf ?}\@warning
       {Citation `\@citeb' on page \thepage \space undefined}}%
\hbox{\csname b@\@citeb\endcsname}}}{#1}}

\catcode`\@=12
\preprint{%
\vspace*{-24 pt}%
\begin{flushright}%
FTUV-12--1121\\
KA-TP--41--2012\\
LPN12--126\\
PSI--PR--12--08\\
SFB--CPP--12--92\\
\end{flushright}
}

\title{$ZZ$+jet production via gluon fusion at the LHC}

\author[a]{Francisco Campanario,}
\author[b,c]{Qiang Li,}
\author[a]{Michael Rauch,}
\author[b]{Michael Spira}

\affiliation[a]{Karlsruhe Institute of Technology,\\Institute for Theoretical Physics, 76128 Karlsruhe, Germany}
\affiliation[b]{Paul Scherrer Institut,\\CH--5232 Villigen PSI,
  Switzerland}
\affiliation[c]{Peking
University,\\State Key Laboratory of Nuclear Physics and Technology,
Beijing, 100871, China}

\emailAdd{francisco.campanario@kit.edu}
\emailAdd{qliphy0@pku.edu.ch}
\emailAdd{michael.rauch@kit.edu}
\emailAdd{Michael.Spira@psi.ch}

\abstract{
Pair production of $Z$ bosons in association with a hard jet is an
important background for Higgs particle or new physics searches at the
LHC.
The loop-induced gluon-fusion process $gg \to ZZg$ contributes formally
only at the next-to-next-to-leading order. 
Nevertheless, it can get enhanced by the large gluon flux at the LHC,
and thus should be taken into account in relevant experimental searches.
We provide the details and results of this calculation, which
involves the manipulation of rank-5 pentagon integrals. Our results show
that the gluon-fusion process can contribute more than $10\%$ to the
next-to-leading order QCD result and increases the overall scale
uncertainty. 
Moreover, interference effects between Higgs and non-Higgs contributions
can become large in phase-space regions where the Higgs is far
off-shell.
}

\date{\Date}

\begin{document} 
\maketitle
\flushbottom

\section{Introduction}
\label{intr}

The Large Hadron Collider (LHC) is presently running with a
center-of-mass energy
of 8 TeV and instantaneous luminosity surpassing already the Fermilab's
Tevatron, with the possibility of being upgraded to the designed value,
i.e., 14 TeV and $10^{34}\;cm^{-2} s^{-1}$, in 2014~\cite{lhc}. The
unprecedented high collision energy and luminosity are necessary for
discovering Higgs particles and new physics beyond the Standard Model
(SM).  However, the higher the collision energy the more complex event
topologies get involved. In particular, hadron collision events with
multiple hard particles and large jet multiplicities deserve a
careful treatment. 

In this paper, we investigate $ZZ+j$ production at the LHC, leading to
the following event topologies
\begin{eqnarray}\label{siga}
pp\rightarrow 4\,\rm{leptons}~+ {\rm jet} + \mathrm{X}\,, \quad
2\,\rm{leptons}~+\dsl{E}_{\mathrm{T}}+ {\rm jet}+ \mathrm{X} \,. 
\end{eqnarray}
This process thus contributes for example as background to Higgs + jet
production with Higgs decays into Z boson pairs, or slepton pair + jet
production with slepton decays into lepton and the lightest
super-symmetric particle (LSP).
The next-to-leading order (NLO) QCD calculation of the corresponding
parton level processes at tree level
\begin{eqnarray} \label{parton1}
q\bar{q} \to ZZg\,,\quad
qg \to ZZq\,,\quad
\bar{q}g \to ZZ\bar{q}\,
\end{eqnarray}
has been performed in Refs.~\cite{ZZj_NLO_partial,Binoth:2009wk} without
subsequent $Z$ decays. The resulting QCD corrections are significant
(e.g., the K factor is about 1.4 for $P_T^{j}>50$\,GeV at the 14 TeV
LHC) and reduce the scale uncertainty significantly.

On the other hand, the IR-safe part of the higher-order QCD corrections,
i.e. the loop-induced gluon-fusion (GF) process 
\begin{eqnarray} \label{parton2}
gg \to ZZg\,,
\end{eqnarray}
which formally is of the next-to-next-to-leading order (NNLO), can be
enhanced by the large gluon flux at the LHC, as already shown in
previous comparative studies~\cite{ggVV} on vector boson ($V$) pair
productions via GF or $q\bar{q}$ collision, i.e. 
\begin{eqnarray} \label{parton3}
gg \to VV\, \quad \rm{vs.} \quad q\bar{q} \to VV\, .
\end{eqnarray} 
A similar study for the $W$ boson case shows contributions of about 10
percent to the next-to-leading order QCD result~\cite{Melia:2012zg}.
Recently, also a study of the GF process appeared, but neglecting the
Higgs contributions~\cite{Agrawal:2012df}.

Moreover, the GF process~(\ref{parton2}) corresponds to part of the real
emission contributions of the NLO QCD corrections to the loop-induced
$Z$ pair production via GF $gg \to ZZ $ and thus may be crucial in
reducing the theoretical uncertainty for a precise measurement of $Z$
boson pair production, and also for the inclusive Higgs search via
$gg\rightarrow H \rightarrow ZZ$~\cite{Cahn:1986pe}. 

Based on the above mentioned motivations, we are reporting in this paper
on the calculation and results of the $2\rightarrow 3$ GF
process~(\ref{parton2}) at the 7, 8 and 14 TeV LHC\footnote{As for the
1.96 TeV Tevatron we have checked that the GF production rates are tiny
($\lesssim$ 1\% of the $q\bar{q}$ NLO QCD ones) due to the small gluon
flux and is thus not discussed here.}. We, therefore, omit GF
production channels with quarks in the initial
and final state since they interfere with the LO
contributions already at NLO~\footnote{The interference effects of
  $gq\to ZZ q$ production via GF with the LO contribution has been
  computed in~\cite{Binoth:2009wk} and are below 1\%}. 
Furthermore, since we are mostly interested in the effects of the
integrated cross section of $ZZ+j$ production in GF, which is not sensitive to the $Z$ decays, in accordance with
Ref.~\cite{Binoth:2009wk}, the leptonic decay of the Z bosons
and off-shell effects, e.g. $\gamma*\to \ell^+\ell^-$, are not considered.
The paper is organized as
follows. In section~\ref{calculations} we describe the calculation. In
section~\ref{results} we present numerical results and their discussion.
Finally we conclude in section~\ref{sec:end}.


\section{Calculation}

\label{calculations}

We have implemented two independent Monte Carlo programs which rely on
different approaches. Cross checks have been performed at the amplitude
level for fixed phase-space points and also for integrated cross
sections, getting agreement at the double-precision level and within 
statistical errors, respectively.

\begin{figure}
\begin{center}
\includegraphics[width=0.67\textwidth]{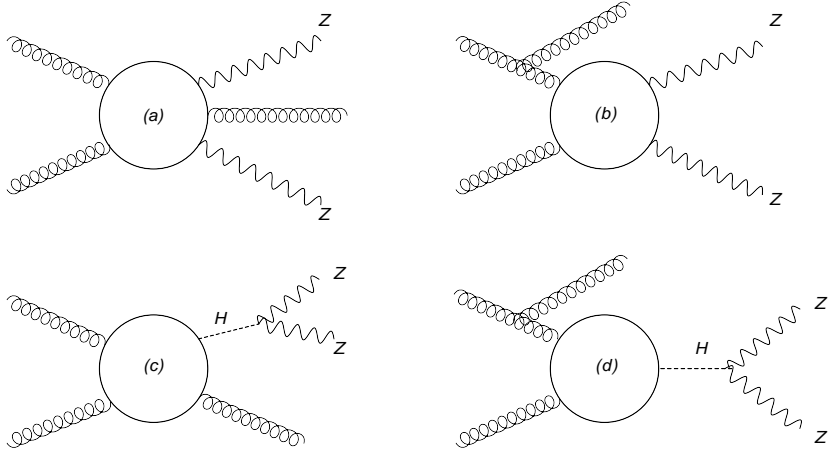}
\end{center}
\caption{Generic Feynman diagrams generated by Jaxodraw~\cite{Jaxodraw}
for the partonic process $gg\rightarrow ZZg$, corresponding to the 4
topological classes. Taking into account all possible permutations, one
gets 12 diagrams for (a), 9 for (b), 3 for (c) and 3 for (d). In
addition, one needs to sum over the fermion types and flow directions
within the fermion loop.}
\label{fd}
\end{figure}

The relevant one-loop Feynman diagrams and amplitudes for the partonic
process $gg\rightarrow ZZg$ are shown in Fig.~\ref{fd}.
The diagrams are grouped in 4 topological classes. The two diagrams in
the upper row correspond to continuum production of the two $Z$ bosons,
either via a pentagon or via a box diagram. The two diagrams in the
bottom row both involve a, possibly virtual, Higgs boson, which then
decays into a $Z$ pair. This is mediated by either box or triangle
graphs. 
The Higgs mass dependence of the latter ones and interference effects
between Higgs and continuum diagrams will be discussed in
Sec.~\ref{results}. 

For program 1, the Feynman amplitudes are generated with FeynArts
3.5~\cite{FeynArts}  and then manipulated with FormCalc
5.3~\cite{FormCalc}\footnote{Note the naive $\gamma_5$ scheme~\cite{ng5}
is employed in FormCalc. The discussion on its validity in practical one
loop calculations in anomaly-free theories can be found e.g.~in
Refs.~\cite{Jegerlehner:2000dz,Dittmaier:2009un}.}. The Fortran
libraries\footnote{The size of the resulting Fortran library for the
helicity amplitude evaluation is about 300 Mb.} generated with FormCalc
are linked with our Monte Carlo integration code for final use. The
tensor integrals are evaluated with the help of the LoopTools-2.5
package \cite{FormCalc,LTEfun}, which employs the reduction method
introduced in Ref.~\cite{PentagonA} for pentagon tensors up to rank 4,
and Passarino-Veltman reduction for the lower point ones up to
boxes~\cite{Passarino:1978jh}.  In our case rank-5 pentagon tensor
integrals are needed in addition, as can be inferred from the fact that
the 5 external particles are all vector bosons. We thus have modified
LoopTools-2.5 to implement the reduction method for pentagon tensor
integrals up to rank 5 as proposed in Ref.~\cite{PentagonB}. Finally,
the resulting regular scalar integrals are evaluated with the FF
package~\cite{FF}. The UV and IR divergent scalar integrals have already
been encoded into this version of LoopTools within dimensional
regularization, which we have explicitly cross-checked against
QCDloop~\cite{QCDloop}.  

Although the reduction procedure in Ref.~\cite{PentagonB} can avoid
inverse Gram determinants of external momenta in the reduction step from
5-point to 4-point integrals, the reduction of lower-point tensor
integrals cannot within the Passarino-Veltman algorithm. The problem is
more severe in our case which involves squared loop amplitudes. 

To improve the numerical stability problem due to vanishing Gram
determinants further modifications have been made. First, we have
implemented in LoopTools the so called `Alternative Passarino-Veltman
reduction' for triangle and box tensor integrals, as introduced in
Ref.~\cite{PentagonB}, which changes the calculating order of tensor
coefficients and results in better numerical convergence behavior than
the conventional Passarino-Veltman reduction.  Second, we have imposed a
jet-measure-like cut to simply cut away a small dangerous region in
phase space to avoid numerical problems. The total contribution of this
region will turn out to be small. 
\begin{eqnarray} \label{ktcut}
&&\min{\bigg(K_T^{i,j}, P_T^i, P_T^j\bigg)} > K_T^{\rm{cut}}, 
\\
&& K_T^{i,j}\equiv \min{\bigg(P_T^i, P_T^j\bigg)} \sqrt{\Delta y_{ij}^2+\Delta\phi_{ij}^2}{\bigg/}0.6\,,
\end{eqnarray}
where $i,j=1,2,3 \, (i\ne j)$ run over the final state particles. Here
$y$ is the rapidity and $\phi$ is the azimuthal angle around the beam
direction. In Sec.~\ref{results}, we will discuss the dependence on the
choice of $K_T^{\rm{cut}}$.

The program 2 is based on the structure of the Monte-Carlo program
\textsc{Vbfnlo}~\cite{Arnold:2008rz}. We use the effective current
approach~\cite{Hagiwara:1985yu}, which allows us
to compute only four master Feynman diagrams, corresponding to the
diagrams appearing in Fig.~\ref{fd}. This calculation is performed with
the in-house framework described in Ref.~\cite{Campanario:2011cs}, which
uses Mathematica~\cite{Wolfram} and
\textsc{FeynCalc}~\cite{Mertig:1990an}.
We use generic vertices split into left- and right-handed components
such that every physically allowed permutation can be constructed by
contracting with the corresponding effective polarization vectors, which
have been multiplied with the electroweak couplings beforehand.
Each of the diagrams is split into a vector and an axial-vector part by
isolating the $\gamma_5$ contributions. This allows to apply Fury's
theorem independently for the vector and axial-vector components
reducing the total number of diagrams to be computed by a factor two.
Additionally, for each of the master diagrams, we build
Ward identities replacing the generic vertices by their corresponding
momenta. This allows us e.g.\ to reduce analytically a pentagon of
 rank five into a difference of two boxes and a remainder pentagon of
rank four
\begin{equation}
{\cal P}^{\mu_1 \ldots \mu_5} p_{i,\mu_i} =
{\cal B}_1^{\mu_1 \ldots \hat{\mu}_i \ldots \mu_5}
- {\cal B}_2^{\mu_1 \ldots \hat{\mu}_i \ldots \mu_5}
+ {\cal P}_\text{rem}^{\mu_1 \ldots \hat{\mu}_i \ldots \mu_5} \,,
\quad i  =1,\ldots,5 \,,
\label{ward}
\end{equation}
where $\hat{\mu}_i$ means that the corresponding vertex has been
replaced by its momentum $p_i$.
The remainder, in this case the pentagon 
${\cal P}_\text{rem}^{\mu_1 \ldots \hat{\mu}_i \ldots \mu_5}$, 
vanishes for massless propagators and purely
vectorial couplings (gluon, photon) for the given contraction.
These simplified analytical expressions are used to control the
numerical accuracy of the code. We compare numerically the values given
by the analytically simplified expressions with the master diagrams,
where the polarization vector has been replaced by the corresponding
four-momentum.
We construct all possible Ward identities for each diagram and physical
permutation, e.g.\ all five different ones for the pentagon ${\cal
P}^{\mu_1\ldots \mu_5}$ before. The deviation is then defined as
absolute value of one minus the ratio between the numerically contracted
and the analytically calculated diagram. Where more than one Ward identity is
possible, we take the largest value.
A point is identified as unstable when this value exceeds a given global
value $\epsilon$. In this case, the complete phase-space point is
rejected and the amplitude set to zero.
The dependence of the cross section on the required accuracy $\epsilon$
will be shown in the numerical analysis in the next section.

To reduce the CPU impact of the calculation of these identities, we
factorize out the part that depends on the effective currents and the
couplings, such that the loop dependent part, including
these identities, is only computed for one helicity combination and
re-used for the other ones. This reduces the time needed for the
additional combinations by about a factor four.
Nevertheless, for the final results we apply random helicity summation
to sample more phase-space points. 
For the numerical evaluation of the tensor integrals, we apply the
Passarino-Veltman approach of Ref.~\cite{Passarino:1978jh} up to boxes,
and for a numerically stable implementation of five-point-coefficients we
use the Denner-Dittmaier scheme laid out in Ref.~\cite{PentagonB} with
the set-up and notation of Ref.~\cite{Campanario:2011cs}.
Color factors have been computed by hand and cross-checked with the
program \textsc{Color}~\cite{Hakkinen:1996bb}.

Additionally, we have implemented a two-layer rescue system for
phase-space points where the Ward identities of Eq.~(\ref{ward}) are not
satisfied. 
In the first step, we calculate the diagram again using dedicated
subroutines for small Gram determinants. These involve the evaluation of
three- and four-point functions up to Rank 11 and 9, respectively,
following the notation of Ref.~\cite{Campanario:2011cs}. 
If at this point the Ward identities are still not satisfied, 
we perform the second step of the rescue system.
Here the scalar integrals and tensor reduction routines are evaluated in
quadruple precision. This requires reconstructing the external momenta
in quadruple precision, so that global energy-momentum conservation is
still fulfilled at the higher numerical accuracy while keeping external 
particles on their mass-shell. This is a crucial step for obtaining 
an improved behavior of the quadruple precision routines. These routines are
only a factor 2-3 slower than the double precision ones, in contrast to the
10-20 factor one would obtain using quadruple precision for the complete
diagram, thus reducing significantly the overall slowing factor of the
rescue system. 
With this system we find the percentage of phase-space points that does not pass 
the Ward identities for a requested accuracy of
$\epsilon=10^{-3}$ is completely negligible, see Table~\ref{Tabl:Ward}. The
additional CPU time required is below 10\% for this accuracy. A detailed discussion of the numerical
impact is postponed to the following section.

Furthermore, in this approach the cut of Eq.~(\ref{ktcut}) is not needed
to obtain stable results since singular points are correctly identified by
the Ward identities. Nevertheless, we have implemented the cut for
comparing with program 1. Final results will be given with program 2
without imposing the $K_T$ cut and activating the two rescue systems
demanding a global accuracy of the Ward identities of $\epsilon=10^{-3}$.

In both programs, we have checked the cancellation of UV and IR
divergences in our calculations. 


\section{Numerical Results}
\label{results}

In this section, we present the integrated cross sections and differential
distributions for $ZZ+j$ production at the LHC with a center-of-mass
energy of 7, 8 and 14 TeV. We impose the following set of cuts
\begin{eqnarray}\label{cut1}
&& |\eta_{j}|< 4.5 \,,\,\qquad P_T^{j}>50\,{\rm GeV}\,, 
\end{eqnarray}
to identify massless partons with jets. 

We set the top quark mass to $m_t=171.3$\,GeV, the bottom quark mass to
$m_b=4.6$\,GeV, and the other light quark masses to zero.  In accordance
with Ref.~\cite{Binoth:2009wk}, we set explicitly
\begin{eqnarray}\label{inputs}
 M_Z = 91.188 \text{ GeV}\,,\quad \alpha(M_Z) = 0.00755391226\,,\quad \sin\theta_W^2=0.222247\,.
\end{eqnarray}
We use a constant Higgs width and the default Higgs mass is chosen to be $M_H=126$\,GeV. 
Results for
$M_H=120,140,\,200$ and $400$\,GeV will also be shown.  The corresponding
Higgs decay widths are obtained by HDECAY~\cite{Djouadi:1997yw}:
\begin{eqnarray}\label{hwidth}
\Gamma_H=0.003485,\,0.003812,\,0.00812,\,\,1.426\,\,\rm{and}\,\,28.89\,\,\rm{GeV}\,,
\end{eqnarray}
for $M_H=120,126,\,140,\,200$ and $400$\,GeV, respectively.  Throughout our
calculation, $ZZ+j$ production rates via GF are calculated with CTEQ6L1
parton distribution functions (PDFs) \cite{Pumplin:2002vw} with the
default strong coupling value $\alpha_s(M_Z)=0.130$ using the
implementation in LHAPDF~\cite{Whalley:2005nh}. Our canonical
choice for the renormalization and factorization scales is
$\mu_R=\mu_F=M_Z$.

\begin{figure}[!ht]
\begin{center}
\includegraphics[width=0.48\textwidth]{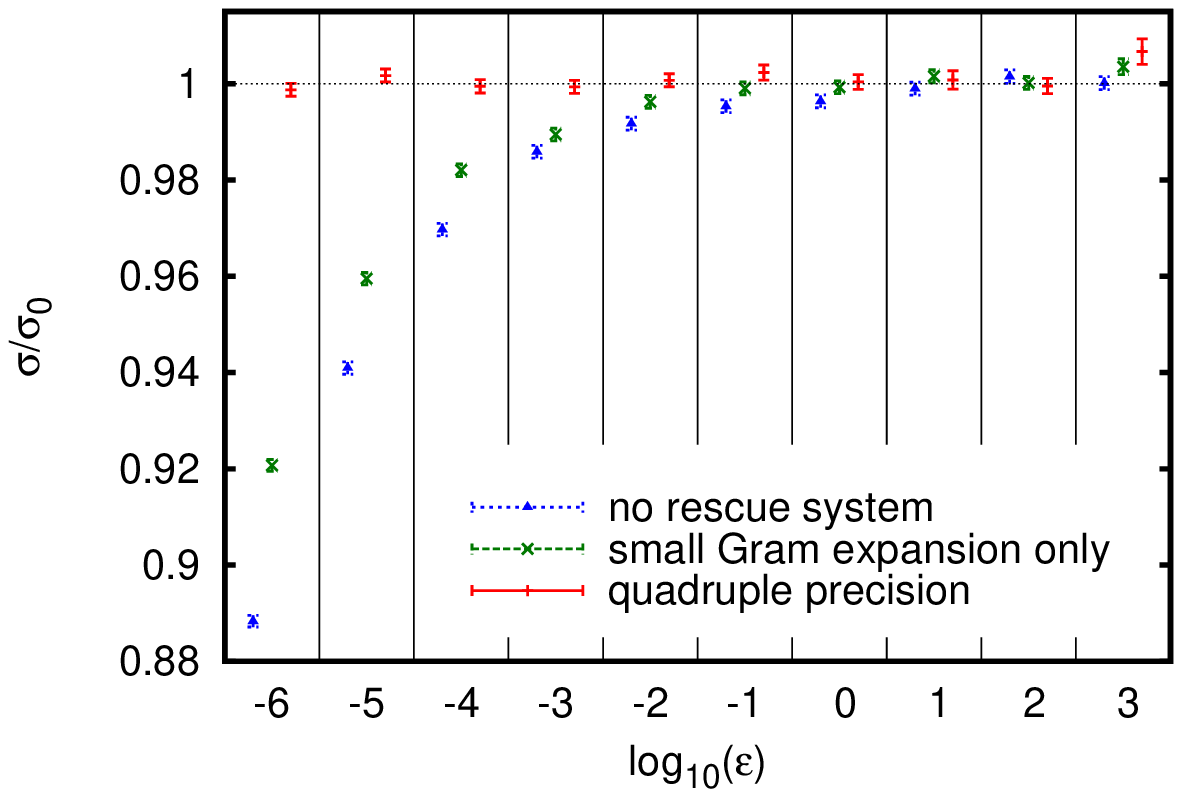}\quad
\includegraphics[width=0.48\textwidth]{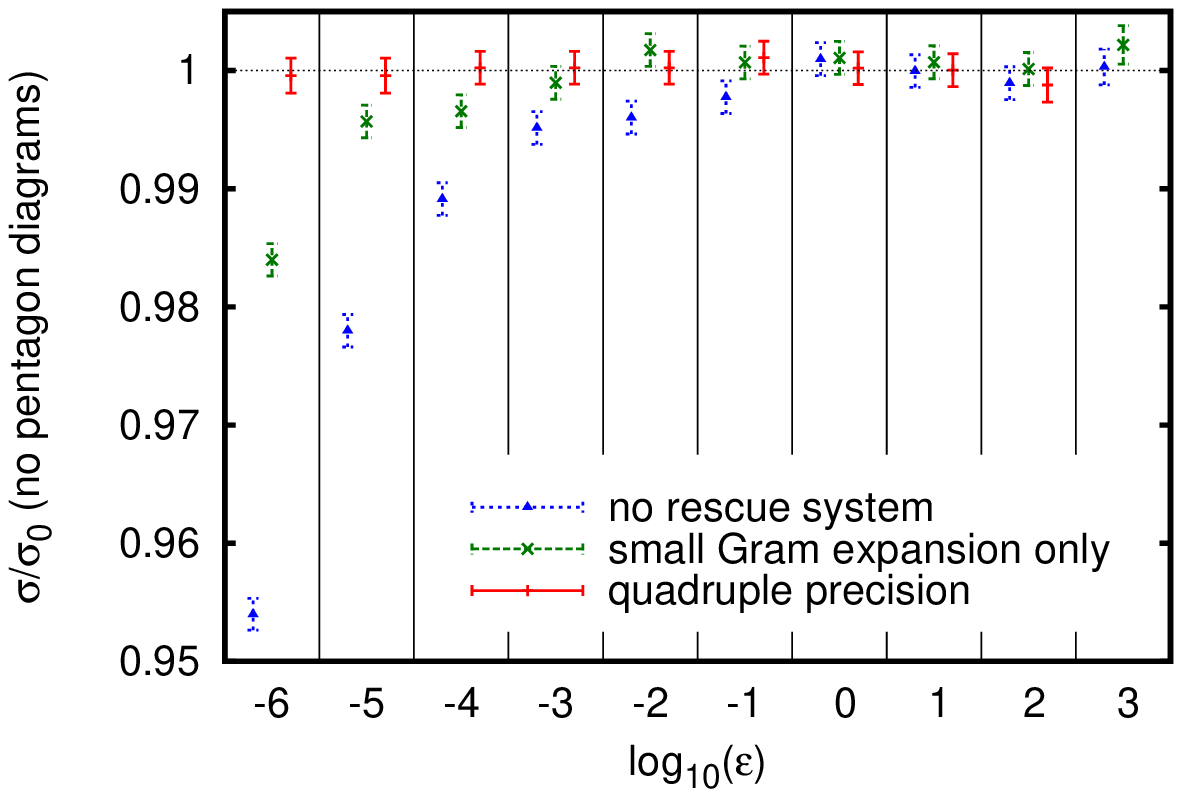}
\vspace*{-2.5em}
\end{center}
\caption{Dependence of the $ZZ+j$ cross section on the value of the
requested Ward identity accuracy $\epsilon$ and the different steps of
the rescue system. The cross sections are normalized to the average
cross section $\sigma_0= 320.8(2)~\text{[fb]}$ of the $\epsilon=10^{-6}$ to $10^{-2}$
quadruple precision runs. Results are generated for the LHC at a
center-of-mass energy of 14 TeV and a Higgs mass of 126 GeV. {\textit
Left:} All diagrams taken into account. {\textit Right:} Only diagrams
up to boxes considered.}
\label{Ward}
\end{figure}

In program 2, we use Ward identities to identify problematic
configurations and use a two-layer rescue system. In
Fig.~\ref{Ward}, we present the cross section for different values of
the demanded accuracy $\epsilon$. We show results without applying any
rescue system as well as the one including only the small Gram
determinant expansion and where both this and quadruple precision for
scalar and tensor integrals have been switched on.
\begin{table}[!ht]
\begin{center}
\begin{tabular}{|l||r@{.}l|l|r@{.}l|l|r@{.}l|l|} 
\hline
\multirow{2}{*}{Accuracy $\epsilon$}& 
\multicolumn{3}{|c|}{before step 1} & 
\multicolumn{3}{|c|}{after step 1} & 
\multicolumn{3}{|c|}{after step 2} 
\\\cline{2-10}
&
\multicolumn{2}{|l|}{
failure rate
} & $\sigma_{\text{MC}}~\text{[fb]}$ & 
\multicolumn{2}{|l|}{
failure rate
} & $\sigma_{\text{MC}}~\text{[fb]}$ &  
\multicolumn{2}{|l|}{
failure rate
} & $\sigma_{\text{MC}}~\text{[fb]}$ 
\\\hline
$10^{-6}$ & 16&8  \% & 285.0(4) &11&1   \%   & 295.3(4)  & 0&036\% &320.4(4)\\                     
$10^{-5}$ &  9&9  \% & 301.8(4)& 5&7   \%    &  307.8(4) & 7&$3\cdot10^{-3}$ \%    &321.3(4)  \\
$10^{-4}$ &  5&7  \% & 311.1(4)& 2&9   \%    &  315.0(4) & 1&$9\cdot10^{-3}$ \%    &320.6(4)  \\ 
$10^{-3}$ &  3&1  \% & 316.2(4)& 1&5   \%    &  317.4(4) & 3&$9\cdot10^{-4}$ \%    &320.6(4)  \\
$10^{-2}$ &  1&7  \% & 318.1(4)& 0&75  \%    &  319.6(4) & 1&$0\cdot10^{-4}$ \%    &321.0(4)  \\
$10^{-1}$ &  0&94 \% & 319.3(4)& 0&39  \%    &  320.5(4) & 1&$7\cdot10^{-5}$ \%    &321.5(4)  \\
$10^{0}$  &  0&54 \% & 319.6(4)& 0&20  \%    &  320.5(4) & \multicolumn{2}{|l|}{0}&320.9(5)  \\
$10^{1}$  &  0&30 \% & 320.5(4)& 0&10  \%    &  321.3(4) & \multicolumn{2}{|l|}{0}&321.0(6)  \\
$10^{2}$  &  0&19 \% & 321.2(5)& 0&048 \%    &  320.8(4) & \multicolumn{2}{|l|}{0}&320.6(5)  \\
$10^{3}$  &  0&12 \% & 320.8(4)& 0&026 \%    &  321.9(5) & \multicolumn{2}{|l|}{0}&322.9(9)  \\\hline
\end{tabular}
\end{center}
\caption{Percentage of unstable points depending on the accuracy of the
  Ward identity test, cf.\ Eq.~(\ref{ward}) and cross section results
  for the given set up. Values are given without
applying any rescue system (left columns), after applying an expansion
for small Gram determinants (step 1, middle columns), and after
calculating the loop integrals for still failing points in quadruple
precision (step 2, right columns). Approximately 6 million phase-space
points have been calculated for each entry.}
\label{Tabl:Ward}
\end{table}

One can see that for an accuracy of the Ward identity test above
$10^{-2}$ the double precision results both with and without
the dedicated tensor integrals for small Gram determinants agree with
the quadruple precision ones better than 1$\%$. 
This agreement is better than one would naively expect if all rejected
points contributed with the same average value to the integrated cross 
section as the accepted ones. In this case, the cross section corrected
for the missing phase-space points can
be estimated as $\sigma_{\text{corr}} =
\frac{\sigma_\text{MC}}{1-R_{\text{failed}}}$, where $R_{\text{failed}}$ and $\sigma_\text{MC}$ 
are the relative rate of failed points and the output of the cross
section obtained with the Monte-Carlo program, respectively, given in Table~\ref{Tabl:Ward}.
With this prescription one obtains corrected cross sections which
increasingly exceed the quadruple precision results as we go to smaller
values of the gauge test parameter $\epsilon$. 
This reflects the fact that the rejected points do not belong to any
particular enhanced region of the phase-space.

On the other hand, it is clearly visible that the rescue system for
small Gram determinants does not solve the problem of the
instabilities. This is due to the fact that for a given
phase space point for which the Ward identity is not satisfied with the
demanded accuracy, there is always some physical permutation that
involves not only small Gram determinants but also small Cayley
determinants $X_{0k}$ and $X_{ij}$, so that the expansion breaks down.
Note, also, that our Ward identity check is very restrictive since we do
not check whether the resulting diagram gives a numerically relevant
contribution to the whole event, i.e, a global gauge check demanding the
same accuracy will probably result into smaller failure rates since
most of the configurations for which the Ward identity is not satisfied
contribute little to the whole event. This can be inferred
seeing the nice convergence to the right result even allowing deviations
of the Ward identities of two orders of magnitude, $\epsilon=10^2$. 

Nevertheless, we have not investigated this further, 
and, instead, to solve the problem, we change to quadruple precision, which is now
also supported in the latest version of \textsc{gfortran} and has a small impact in
terms of CPU time in our program. After this step, the instabilities fall well below the per mille level for an
accuracy of the Ward identity gauge test of $10^{-6}$ despite the fact
of only using quadruple precision for the scalar and tensor integrals. The
small fraction of failed points left after this step points to the fact
that the loss of precision is really due to the presence of small Gram
determinants. 
Note also that when allowing large deviations of the Ward identities of
three orders of magnitude, $\epsilon=10^{3}$, some of the runs
already start to feel the instabilities, which translates into
large weights yielding larger cross sections and statistical
errors. Without applying the Ward identities, $\epsilon \rightarrow \infty$,
most of the runs with different random number seeds simply
give arbitrarily huge values with corresponding huge error bars, if not
initialized with an optimized grid.
In the following, we stick to the set up to switch on the
rescue system for an accuracy of the Ward identity of
$\epsilon=10^{-3}$. The CPU cost is at the 10$\%$ level even with the
poor success rate of the first step, which we keep for academic reasons.
\begin{figure}[!ht]
\begin{center}
\includegraphics[width=0.6\textwidth]{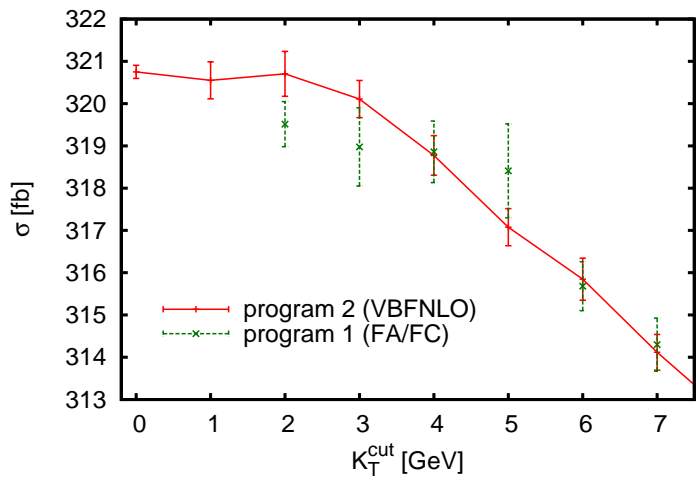}
\vspace*{-2.5em}
\end{center}
\caption{$K_T^{\rm{cut}}$ dependence of $ZZ+j$ production rates via GF
  at the 14 TeV LHC. The statistical error bars for both programs are also
  shown. For cut values smaller than about 3 GeV the effect of the cut
  is below the integration error.}
\label{kcut}
\end{figure}

As mentioned in Sec.~\ref{calculations}, for program 1 we employ 
$K_T^{\rm{cut}}$ for simplicity to avoid the problem of numerical
instability from vanishing Gram determinants. We show in Fig.~\ref{kcut}
the $K_T^{\rm{cut}}$ dependence of the integrated cross section for $ZZ+j$
production via GF with the cuts of Eq.~(\ref{cut1}) at the 14 TeV LHC for a Higgs mass of
$M_H=\,140$\,GeV. For both programs the speed of the
code is about 4 CPU days on a state-of-the-art computer with the
statistical error on the Monte Carlo integration better than 2 per
mille. In the region
of $K_T^{\rm{cut}} \lesssim 3$\,GeV, all the points agree well with each
other within statistical errors, which shows that the $K_T^{\rm{cut}}$
dependence is small. However, for  $K_T^{\rm{cut}}$ smaller $\lesssim
5$\,GeV, the statistical errors on program 1 are hard to improve
further, as expected, and the percentage of runs that gives nonsensical
results using different initial seeds increases for decreasing values of
the $K_T^{\rm{cut}}$ cut.  With increasing
$K_T^{\rm{cut}} \gtrsim 4$\,GeV, deviations from the small
$K_T^{\rm{cut}}$ limit start to show up and become apparent as the phase
space is reduced more and more. In the following, we use program 2
without applying any $K_T^{\rm{cut}}$ for the numerical results.

\begin{table}
\centering{
\begin{tabular}{|c|c|c|c|c|c|} \hline
7 TeV LHC / $M_H$       & $120$\,GeV & $126$\,GeV & $140$\,GeV & $200$\,GeV & $400$\,GeV \\ \hline
$\sigma(\text{cont})$   & \multicolumn{5}{|c|}{53.08(8)} \\ \hline
$\sigma(H)$             & 
  $6.081(3)$ & $6.255(4)$ & $6.753(4)$ & $173.11(6)$ & $111.03(7)$ \\ \hline
$\sigma(\text{cont}+H)$ & 
  $48.93(7)$ & $48.74(7)$ & $48.71(7)$ & $224.2(3)$ & $160.6(4)$ \\ \hline
$\frac{\sigma(\text{cont}+H)}{\sigma(\text{cont})+\sigma(H)}$ & 
$0.83$ & $0.82$ & $0.81$ & $0.99$ & $0.98$  \\  \hline
\end{tabular}
\caption{Dependence of the $ZZ+j$ production rate via GF (in fb) on
$M_H$ at the 7 TeV LHC and the interference effects between Higgs and
non-Higgs contributions.}
\label{mh7}}
\end{table}
\begin{table}
\centering{
\begin{tabular}{|c|c|c|c|c|c|} \hline
8 TeV LHC / $M_H$       & $120$\,GeV & $126$\,GeV & $140$\,GeV & $200$\,GeV & $400$\,GeV \\ \hline
$\sigma(\text{cont})$   & \multicolumn{5}{|c|}{79.1(1)} \\ \hline
$\sigma(H)$             & 
  $9.628(6)$ & $9.891(6)$ & $10.642(6)$ & $251.18(8)$ & $172.3(1)$ \\ \hline
$\sigma(\text{cont}+H)$ & 
  $72.3(1)$ & $72.4(1)$ & $72.4(1)$ & $325.1(5)$ & $245.3(6)$ \\ \hline
$\frac{\sigma(\text{cont}+H)}{\sigma(\text{cont})+\sigma(H)}$ & 
$0.81$ & $0.81$ & $0.81$ & $0.98$ & $0.98$  \\  \hline
\end{tabular}
\caption{Same as Table~\ref{mh7}, but for the 8 TeV LHC.}
\label{mh8}}
\end{table}
\begin{table}
\centering{
\begin{tabular}{|c|c|c|c|c|c|} \hline
14 TeV LHC / $M_H$       & $120$\,GeV & $126$\,GeV & $140$\,GeV & $200$\,GeV & $400$\,GeV \\ \hline
$\sigma(\text{cont})$   & \multicolumn{5}{|c|}{354.5(2)} \\ \hline
$\sigma(H)$             & 
  $54.04(3)$ & $55.21(1)$ & $58.65(3)$ & $1034.6(4)$ & $877.8(5)$ \\ \hline
$\sigma(\text{cont}+H)$ & 
  $321.2(2)$ & $320.8(2)$ & $318.1(1)$ & $1342(2)$ & $1184(3)$ \\ \hline
$\frac{\sigma(\text{cont}+H)}{\sigma(\text{cont})+\sigma(H)}$ & 
$0.79$ & $0.78$ & $0.77$ & $0.97$ & $0.96$  \\  \hline
\end{tabular}
\caption{Same as Table~\ref{mh7}, but for the 14 TeV LHC.}
\label{mh14}}
\end{table}

In Tables~\ref{mh7}, \ref{mh8} and~\ref{mh14}, we present $ZZ+j$ production rates via
GF with on-shell $Z$ bosons at the 7 TeV, 8 TeV and 14 TeV LHC, for
$M_H$=120, 126, 140, 200 and 400 GeV, respectively. The results for
Higgs masses different from 126 GeV are included to demonstrate the
effect of interference effects for these masses. While we know from the
experimental searches~\cite{ATLASHiggs,CMSHiggs} that a Higgs boson with
SM coupling strength is not possible there, one with reduced couplings,
\eg{} the CP-even partner in a Two-Higgs-doublet model, is still viable.

$\sigma(\text{cont}+H)$ gives the integrated cross sections including both
Higgs signal ($\sigma(H)$) and continuum ($\sigma(\text{cont})$)
contributions. For $M_H\lsim 2M_Z$, the Higgs contributions are small
compared to continuum production, as the intermediate Higgs is far
off-shell. Note that we only consider production of on-shell $Z$ bosons,
which is the dominant region for the continuum contribution, where we
will focus on in the current work. 
We see that for Higgs masses below the threshold, which includes the
actually observed mass, there is a strong destructive interference
between the continuum and the Higgs diagrams. This leads to integrated cross
sections which are even below the continuum-only result. Compared to the
naive sum, the full result is reduced by roughly 20\%. In contrast for
Higgs masses above the threshold, interference effects are small
and reach at most 4\%.

\begin{figure}
\begin{center}
\includegraphics[width=0.6\textwidth]{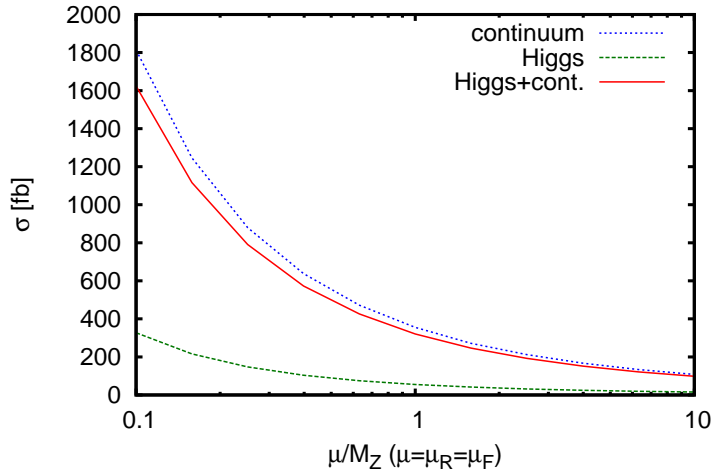}
\vspace*{-2.5em}
\end{center}
\caption{Scale dependence of the integrated cross sections for $ZZ+j$
production at the 14 TeV LHC.}
\label{scale}
\end{figure}
Fig.~\ref{scale} shows the dependence on the renormalization and
factorization scales ($\mu=\mu_R=\mu_F$) of the gluon-fusion $ZZ+j$
production rates at the 14 TeV LHC. 
The scale dependence is rather large. Varying the scale by a factor 2
downwards (upwards), the cross sections change by 54.2\% (-32.5\%),
59.4\% (-34.4\%) and 54.2\% (-32.4\%) for continuum diagrams, Higgs
diagrams and the full result, respectively.
Comparing these numbers with the NLO results for the continuum calculated in
Ref.~\cite{Binoth:2009wk}, the continuum gluon-fusion result gives an
additional 9.7\% contribution to the cross section at the central scale
$\mu=M_Z$ (13.2\% compared to the LO cross section). This changes to 
13.9\% (18.0\%) and 7.0\% (10.0\%) for decreasing and increasing the scale by a
factor two, respectively. The total scale uncertainty of the cross section
increases from +8\% (+13\%) and -6\% (-11\%) for the NLO QCD (LO) result to
+12.1\% (+17.8\%) and -8.4\% (-13.5\%) for NLO+GF (LO+GF), again varying
the scale by a factor two down- and upwards, respectively.

\begin{figure}
\includegraphics[width=0.45\textwidth]{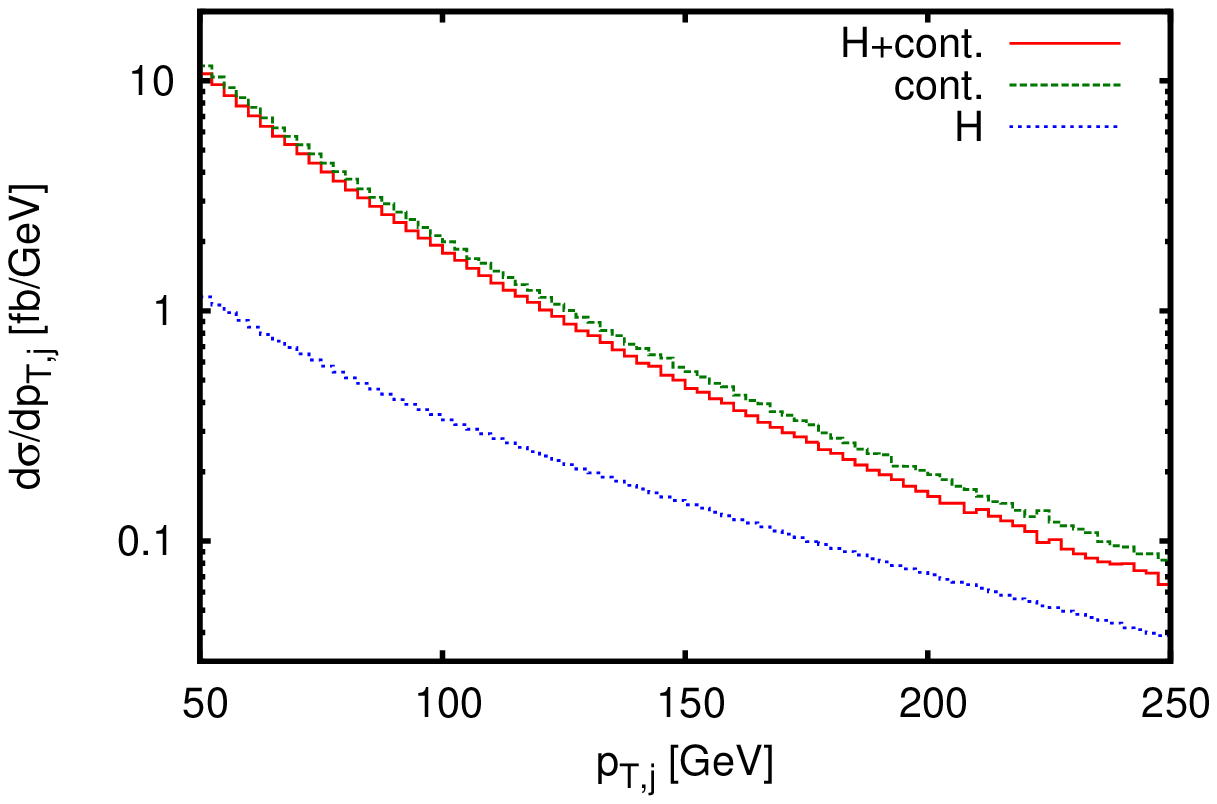}\qquad
\includegraphics[width=0.45\textwidth]{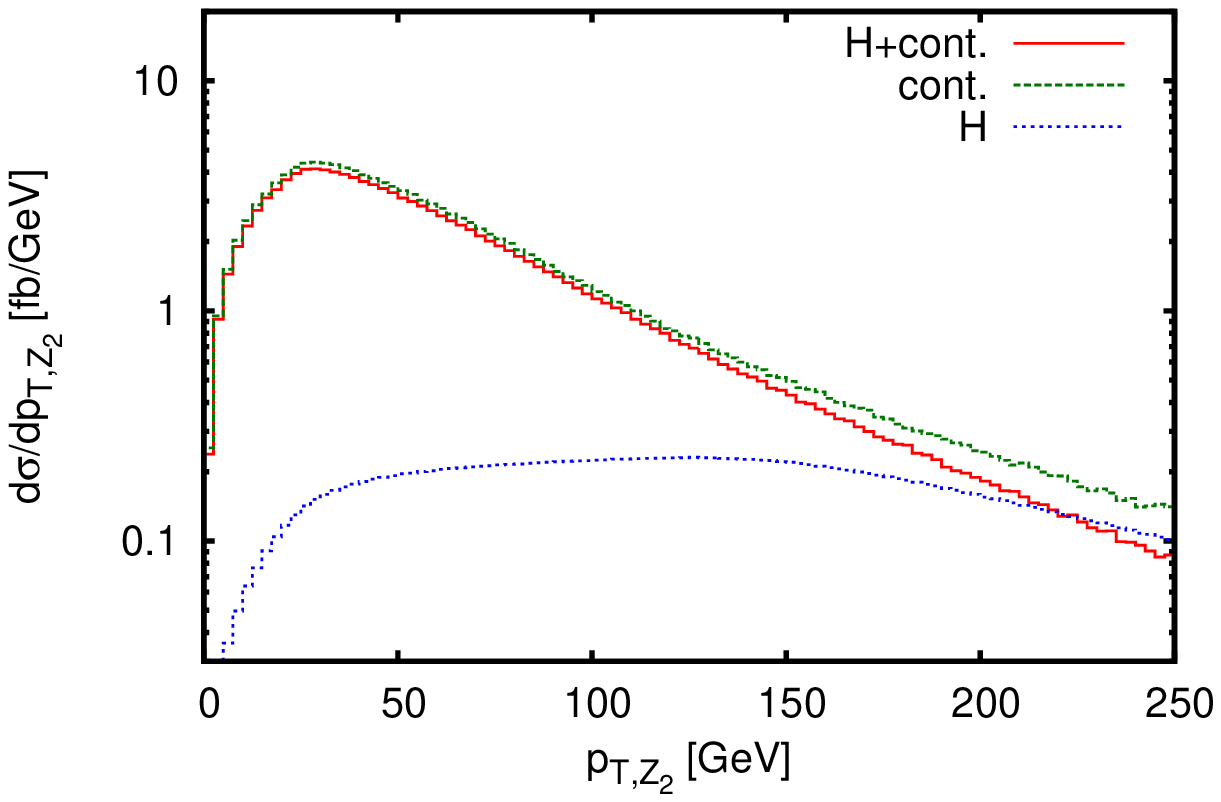}
\caption{Differential cross sections for the $p_T$-distribution of the
jet (\textit{left}) and the Z boson with the smaller $p_T$
(\textit{right}) for the 14 TeV LHC using $M_H$ = 126 GeV. The
individual curves show the contribution of only continuum diagrams
(dashed green lines), only Higgs diagrams (dotted blue) and both types
including interferences (solid red).}
\label{pt}
\end{figure}

In Fig.~\ref{pt} we show, in the left-hand panel, the differential cross
section for the distribution of the transverse momentum of the jet
$p_{T,j}$. Both diagram types lead to increasingly lower cross sections
as we go to higher transverse momenta, but the fall-off of the Higgs
part is less steep than the continuum part. Compared to the tree-level
process~\cite{Binoth:2009wk}, the jet distribution here in gluon-fusion
is softer due to the dilution effects of the fermion loop.  This
behavior is consistent with previous findings in Ref.~\cite{Li:2010fu}
on Higgs+jet production.  On the right-hand side of Fig.~\ref{pt}, we
present the transverse-momentum distribution of the second Z boson
with the smaller $p_T$ value. A similar behavior as for the
jet-$p_T$ is observed, and the cross section for the Higgs-only diagrams
stays almost constant over a large part of the shown range. For
transverse momenta larger than about 200 GeV, it reaches the same order
of magnitude as the continuum contribution and even exceeds the combined
result.  

\begin{figure}
{\centering
\includegraphics[width=0.45\textwidth]{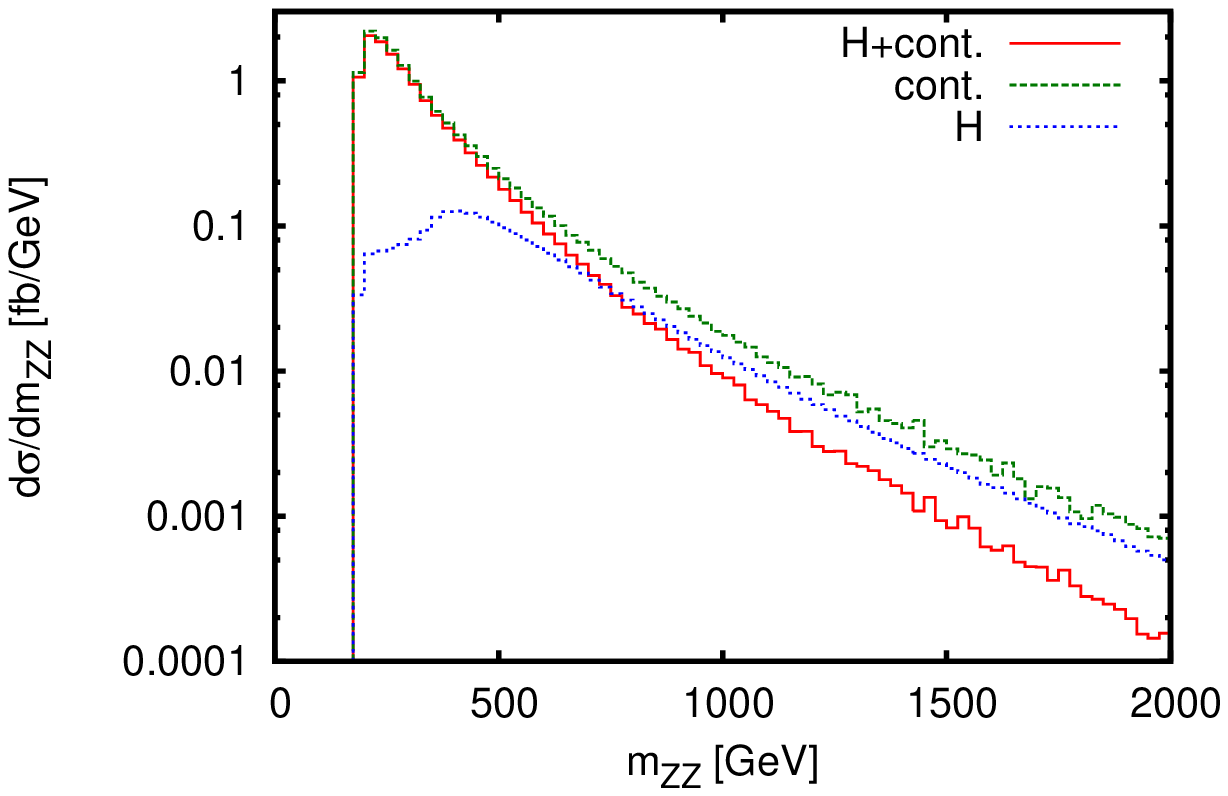}\qquad
\includegraphics[width=0.45\textwidth]{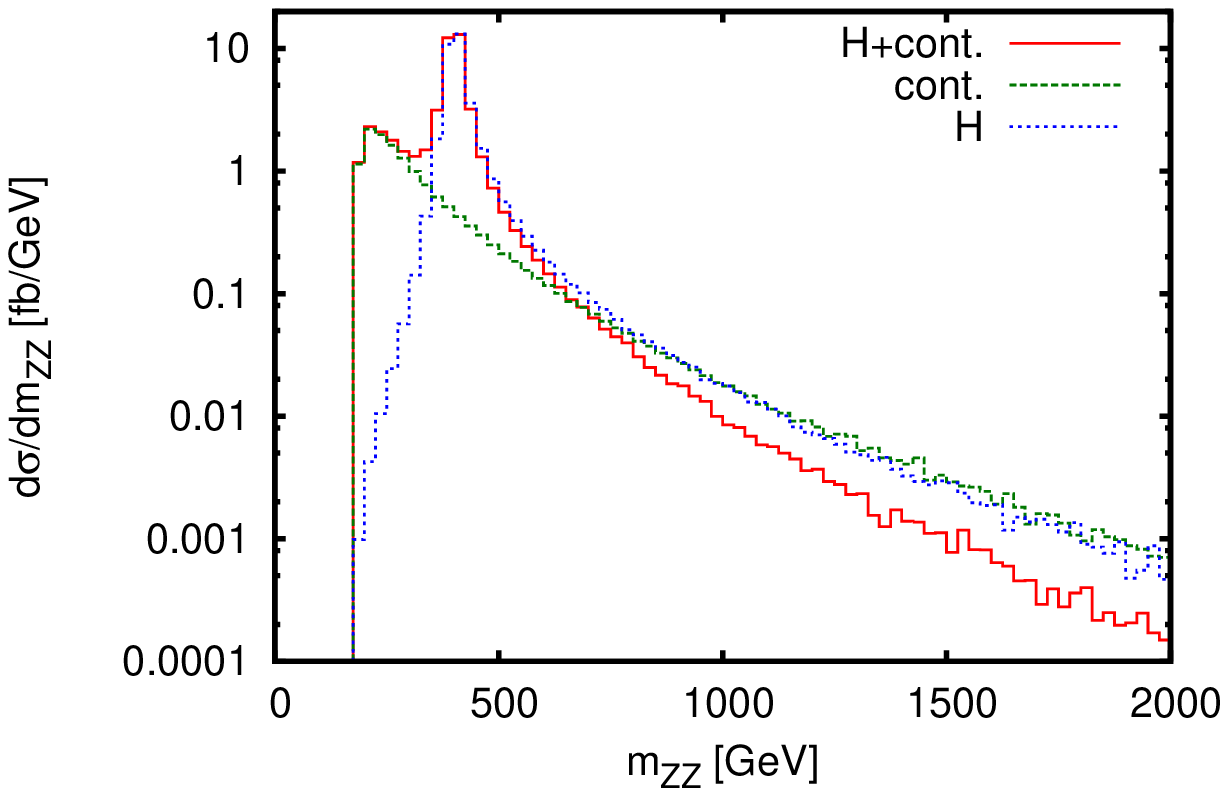}
}
\caption{Differential cross section for the invariant mass of the two
$Z$ bosons for the 14 TeV LHC using $M_H$ = 126 GeV (\textit{left}) and
$M_H$ = 400 GeV (\textit{right}). The
individual curves show the contribution of only continuum diagrams
(dashed green lines), only Higgs diagrams (dotted blue) and both types
including interferences (solid red).}
\label{mzz}
\end{figure}

The origin of this behavior can be understood by looking at the
invariant mass of the two $Z$ bosons shown in Fig.~\ref{mzz}.  For $M_H$
= 126 GeV in the left-hand panel, the continuum diagrams show a peak
directly above the threshold and a fall-off for larger invariant masses.
These are dominated by the loop diagrams with massless quarks of the
first two generations running in the loop. For the Higgs diagrams in
contrast the top-quark loop dominates. Here the peak of the cross
section is just above crossing the $2 m_t$ threshold owing to the P-wave
suppression at threshold due to the CP-even nature of the Higgs boson.
The Z bosons coming from virtual Higgs decays have larger momenta on
average, leading to a harder $p_T$ spectrum.  On the right-hand side we
show in comparison the same plot, but now setting the Higgs mass to 400
GeV. The Higgs resonance in the ZZ invariant mass spectrum is clearly
visible. While for mass values smaller than that constructive
interference between the two diagram types appears, it becomes
destructive for larger values. This is in particular visible at the
large-mass end of the plot, where both contributions have similar size
and the sum of the two is about a factor three smaller. Such a behavior
has already been observed and explained in $gg\rightarrow ZZ$
production~\cite{ggZZ}. For large momenta the longitudinal polarizations
of the $Z$ bosons dominate, which for the continuum couple predominantly
to the top-quark loop, as does the Higgs.  For colliding two on-shell
top quarks, unitarity restoration then immediately requires that for the
longitudinal polarizations continuum $t$- and $u$-channel diagrams and
the $s$-channel Higgs diagram cancel at large invariant masses. This
behavior is unchanged when closing the loop and integrating over the
loop momentum.

\begin{figure}
{\centering
\includegraphics[width=0.45\textwidth]{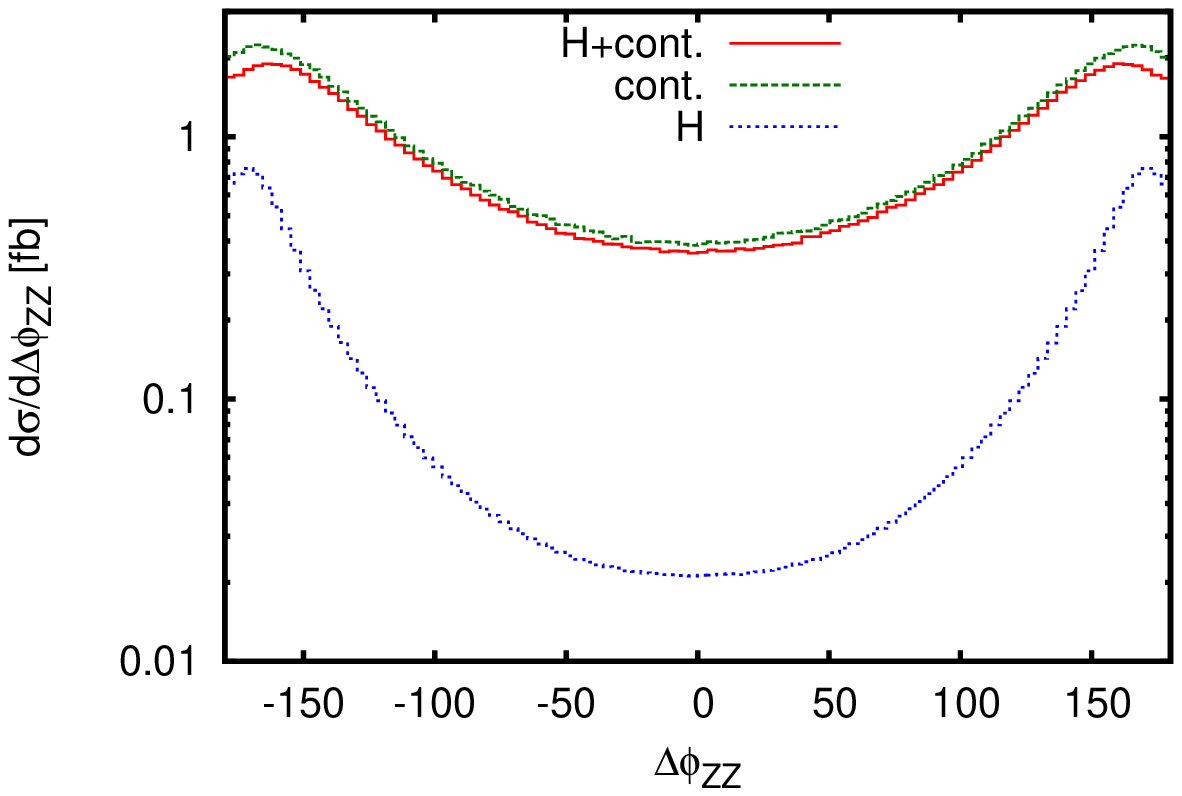}\qquad
\includegraphics[width=0.45\textwidth]{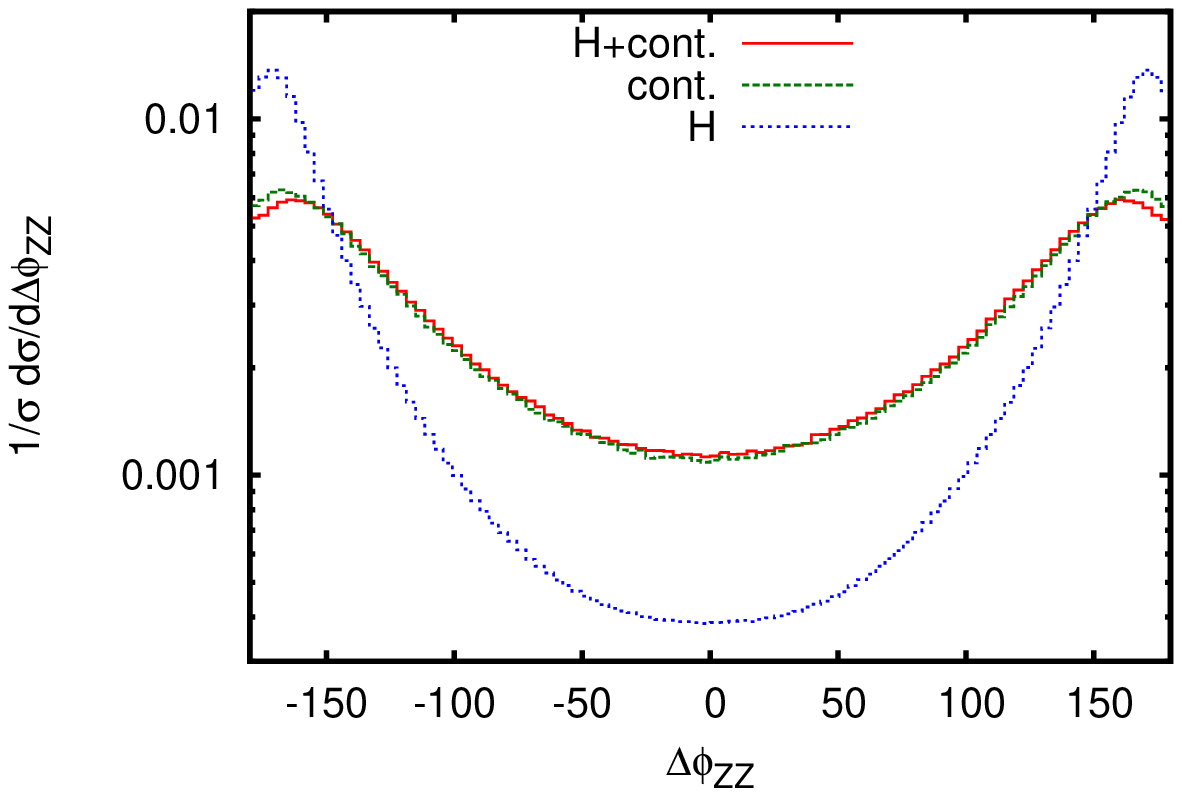}
}
\caption{Differential cross section for the invariant mass of the two
$Z$ bosons for the 14 TeV LHC using $M_H$ = 126 GeV. The
individual curves show the contribution of only continuum diagrams
(dashed green lines), only Higgs diagrams (dotted blue) and both types
including interferences (solid red) for absolute (\textit{left}) and
relative cross sections (\textit{right}).}
\label{deltaphizz}
\end{figure}

Finally, we display the azimuthal angle separation between the two $Z$
bosons in Fig.~\ref{deltaphizz}. Again, we see a significant difference
between the continuum and the Higgs diagrams. In both cases, the two Z
bosons are preferably emitted back-to-back, but the effect is more
pronounced for the Higgs contribution. This is particularly visible in the
right-hand panel, where the differential cross sections are individually
normalized to the integrated one.


\section{Summary}
\label{sec:end}

We have presented a calculation of the loop-induced gluon-fusion
process of $ZZ+j$ production at the LHC. Special attention has been
paid to the numerical problem of vanishing Gram determinants to obtain
stable results. 

We have studied distributions of the final-state particles. Here, the
contribution of the Higgs diagrams develops larger transverse momenta of
the final-state particles. This is due to the crossing of the top-quark
pair production threshold, which increases the production for $ZZ$
invariant masses above this value. Also, for invariant masses larger
than the Higgs mass, destructive interference between Higgs and
continuum diagrams appears, leading to a huge reduction of the
differential cross section.  The effect on the integrated cross section,
however, is about 20\% for a Higgs mass of 126 GeV and, when compared
with the tree-level cross section of $q\bar q$-induced ZZj production,
small.

Additionally, when compared with the known NLO QCD result, the
gluon-fusion part can contribute more than 10\% of the NLO QCD cross
section, especially at small scales $\mu$ and jet transverse momenta
$p_{T,j}$.  Moreover, the GF results increase the scale uncertainties of
the integrated $ZZj$ cross section. For \eg{} the benchmark point in
Fig.~\ref{scale}, the scale dependence is increased by about a factor
1.5. 


\acknowledgments

This work is supported by the European Community's Marie-Curie Research
Training Network HEPTOOLS under contract MRTN-CT-2006-035505 and the
National Natural Science Foundation of China, under Grants No. 11205008.
FC and MR acknowledge support by the Deutsche
Forschungsgemeinschaft via the Sonderforschungsbereich/Transregio
SFB/TR-9 ``Computational Particle Physics'',  the Initiative and
Networking Fund of the Helmholtz Association, contract HA-101(``Physics at
the Terascale''), and by the FEDER and Spanish MICINN under
grant FPA2008-02878.


\appendix


\end{document}